\documentclass[a4paper,10pt,twocolumn]{article}
\pdfoutput=1
\usepackage[T1]{fontenc}
\usepackage[margin=1in]{geometry}
\author{Dorai Ashok S.A. <dash.r@mpipe.net>\\
\textit{\smash{$0^{th}$} Root Software Research Private Limited}\\
\textit{No.87A, Vembuliamman Koil Street, Periya Nolambur, Chennai-600095, India}\\
\textit{Phone: +91 44 2653 2984}}
\title{Secure Shell (SSH): Public Key Authentication over Hypertext Transfer Protocol (HTTP)}
\begin{document}
\maketitle
\begin{abstract}
The Secure Shell (SSH) protocol requires all implementations to support public key authentication method (``publickey'') for authentication purposes \cite{rfc4252}. Hypertext Transfer Protocol (HTTP) applications which provide a SSH client over the web browser need to support ``publickey''. However, restrictions in HTTP, such as Same Origin Policy, make it difficult to perform such authentications. In this document, a system to perform ``publickey'' authentication over HTTP is provided. It is ensured that no compromise is made that would pose a security risk to SSH protocol.
\end{abstract}
\section{Introduction}
The Secure Shell (SSH) protocol is a protocol for secure remote login to servers. It computes a shared key through key exchange mechanisms, such as Diffie-Hellman key exchange \cite{rfc4253}, and establishes a secure channel for communication to servers. Users are authenticated through the secure channel using one of the accepted authentication methods. Commonly used authentication methods include ``password'' and ``publickey''. As the name suggests, in case of ``password'' method, the user is simply required to provide his/her \emph{password} for authentication. However, ``publickey'' method uses public/private key pairs for authentication.

The Hypertext Transfer Protocol (HTTP) is an application-level protocol for data transfer across the internet. It is generic, stateless and can be used for purposes beyond hypertext.
\section{Authentication ``publickey''}
In ``publickey'' authentication method, User creates a public/private key pair and adds the public key on the server. For successful authentication, the SSH protocol requires the \emph{message} \cite{rfc4252}, shown below, to be signed by the private key of the user. And, the signed message should be verified by the server using the public key added by the user. On successful verification, authentication is successful and the user is permitted to login.
\label{sshmsg}
\begin{verbatim}
      string    session identifier
      byte      SSH_MSG_USERAUTH_REQUEST
      string    user name
      string    service name
      string    "publickey"
      boolean   TRUE
      string    public key algorithm name
      string    public key blob
\end{verbatim}
The \verb+session identifier+, shown above, is a \emph{secret} and is specific to that SSH session.

It is common for private keys to be kept in an encrypted format for security reasons. However, this creates the hassle of decrypting the private key for authentication every single time. Authentication agents, such as \emph{ssh-agent} \cite{sshagent}, handle this by managing the decrypted private keys and signing messages for SSH clients on demand.
\section{Authentication agents}
\emph{ssh-agent} implements OpenSSH agent protocol \cite{sshagent} to manage private keys and to \emph{sign} messages. Clients connect to the agent, typically, over an UNIX domain socket to send requests and receive responses. UNIX domain socket is a special file in the file system. To restrict access, the special file has read/write permissions only for the user.

If we could implement an authentication agent, similar to \emph{ssh-agent}, that would manage private keys but accept signing requests over HTTP. That would be an agent which can enable SSH ``publickey'' authentications through HTTP clients.
\section{SSH Web Agent}
Here, we are defining a new authentication agent, \emph{ssh-webagent}, which responds to authentication requests over HTTP. It accepts requests only from user and only on \emph{localhost}. And, due to same origin policy, HTTP clients will not accept content from \emph{localhost} unless Cross Origin Resource Sharing (CORS) \cite{cors} is enabled. It is also recommended that \emph{ssh-webagent} accept HTTP requests over TLS \cite{https} to avoid Mixed Content \cite{mixedcontent}.
\subsection{Trusted Servers}
On receiving an authentication request, it is important to identify the source of the request. A HTTP client could connect to \emph{ssh-webagent} out of any web page, and without verification, a serious security risk exists. To verify the source, a combination of HTTP \emph{Referer} header and public key of the server should be used. While the \emph{Referer} header can be used to identify the public key of the server, the actual verification is successful only if the signature provided by the server is valid.

Server's public key and HTTP Referer changes for each web application. It is therefore recommended that a \emph{trusted servers} file is used to hold the public key of trusted servers along with the expected values of HTTP Referer header. The \emph{trusted servers} file should be kept safe to avoid unauthorized modifications. A sample format for \emph{trusted servers} file has been provided in Appendix \ref{trusted-servers-file-format}.
\subsection{Identifiers}
HTTP is stateless and each request is independent of the other. Each HTTP request has a corresponding response but the order of HTTP requests and state, if any, shall be managed by the application.

In case of authentication over a TCP connection, a session is established and can be uniquely identified by \emph{5-tuple}\footnote{source IP address, source port, destination IP address, destination port, transport protocol}. However, HTTP provides no session and hence \emph{ssh-webagent} has to establish a logical session to the trusted server and assign a unique \emph{identifier}. \emph{ssh-webagent} can interact with multiple trusted servers at the same time. And, \emph{ssh-webagent} shall choose the unique \emph{identifier}.

The state of the authentication process shall be maintained in association with the \emph{identifier}. However, it should be noted that, a failure in the authentication process may not always be known to \emph{ssh-webagent}. So, it is essential to maintain a reasonably short timeout for the authentication process, beyond which the \emph{identifier} and its associated state is destroyed.
\subsection{Local User and Local Host}
Since \emph{ssh-webagent} accepts authentication request on \emph{localhost}, it shall be ensured that the request is from the same user who runs \emph{ssh-webagent}. Support for this feature varies across operating systems and Appendix \ref{connection-owner} shows how the owner of a TCP connection can be identified on systems running Linux kernel.

When \emph{ssh-webagent} binds to a specific port on \emph{localhost}, no other user will be able to use the same port. So, applications which intend to support multiple local users, can choose multiple ports and send authentication request to all of them. A standard has been proposed to use a specific local IP address and port for \emph{ssh-webagent} in Appendix \ref{local-ip-port}.
\subsection{Session}
\emph{ssh-webagent} shall establish a secure session with the trusted server before processing authentication requests. The session is initiated by a signed request from the trusted server with Diffie-Hellman (DH)\footnote{Diffie-Hellman Key Exchange algorithm\cite{rfc4253}} parameters. On successful verification, \emph{ssh-webagent} shall send a Diffie-Hellman response with an encrypted \emph{identifier}. The secret established through Diffie-Hellman key exchange shall be used in the encryption of the \emph{identifier}. And, once the trusted server successfully decrypts this \emph{identifier}, the session is established.

The authentication process following session creation, will use the \emph{identifier} in both clear and encrypted data. \emph{ssh-webagent} shall identify the session using the clear text \emph{identifier}. However, it shall verify that the \emph{identifier} in encrypted data matches the one in clear text.
\section{Proposed Protocol}
In this section, we define a protocol over HTTP that can be used for ``publickey'' authentication. It involves both session creation and authentication process. We start with the format of HTTP requests and responses before discussing each of them in detail.
\subsection{HTTP Request/Response}
\label{http-request-response}
The protocol uses base64 encoded \emph{messages} in both HTTP request and response. HTTP request shall use \emph{method} \verb+POST+ and content type \verb+application/x-www-form-urlencoded+, while HTTP response shall use content type \verb+text/plain+. HTTP request in this format will classify as simple cross-site request under CORS specification \cite{cors}.
\begin{quote}
\verb+P=[Message]&U=[User]&S=[Service]+
\end{quote}

As a \verb+POST+ request, the key \textbf{P} shall be used to pass messages and, optionally, the keys \textbf{U} and \textbf{S} shall be used to pass user and service names\footnote{user and service names are part of SSH message, See Section \ref{sshmsg}} respectively.

\begin{quote}
\verb+[Message]+
\end{quote}

The response body simply contains the message in response to the request and nothing more.
\subsection{Message}
\label{message}
Message is the basic unit of the protocol. A single \emph{message} is received in the HTTP request and a single \emph{message} is sent in the HTTP response. Message uses a binary format and data types used, in representation of the format, are as defined in RFC 4251 \cite{rfc4251}, Section 5 ``Data Type Representations Used in the SSH Protocols''.
\begin{quote}
\begin{tabular}{l p{1cm} l}
string &  & ``SSHWebAgent''\\
byte &  & version\\
byte &  & type\\
string &  & data\\
\end{tabular}
\end{quote}
\begin{description}
\item[version] shall indicate the version of the protocol and shall be used to check compatability of \emph{ssh-webagent} and \emph{trusted server}.

\begin{tabular}{l p{1cm} l}
VERSION\_1\_1 && 0x11\\
\end{tabular}
\item[type] shall indicate the type of data the message holds.

\begin{tabular}{l p{1cm} l}
KEX\_DH\_REQUEST && 0x02 \\
KEX\_DH\_RESPONSE && 0x03 \\
PRIVATE && 0x04 \\
\end{tabular}
\item[data] holds request parameters and response values in binary format. The format varies based on \emph{type} and, hence, are described in their respective sections.
\end{description}
\subsection{Session}
Session is initiated by a message from \emph{trusted server} containing key exchange parameters and its signature. The message will contain \emph{data} of type KEX\_DH\_REQUEST in the format described below,
\begin{quote}
\begin{tabular}{l p{1cm} l}
mpint &  & p\\
mpint &  & g\\
mpint &  & e\\
string &  & d\\
string &  & k\\
string &  & sign\\
\end{tabular}
\end{quote}
\begin{description}
\item[p, g] are Diffie-Hellman key exchange parameters prime number and generator, respectively.
\item[e] is the computed value of \emph{trusted server} in key exchange
\item[d] is session data, if any, added by \emph{trusted server}. It shall be used by \emph{ssh-webagent} without any interpretation.
\item[k, sign] are the public key and signature, of the \emph{trusted server}, represented in their respective formats as defined in RFC 4253 \cite{rfc4253}, Section 6.6 ``Public Key Algorithms''.
\end{description}
\subsubsection{Verification}
On receiving the message, \emph{ssh-webagent} uses the trusted servers file and checks if the public key and the HTTP Referer are trusted. On a successful match, the message shall be checked for authenticity through signature verification. The following values are used in the specified order for signing and verification,
\begin{quote}
\begin{tabular}{l p{1cm} l}
mpint &  & p\\
mpint &  & g\\
mpint &  & e\\
string &  & method\\
string &  & referer\\
string &  & k\\
string &  & d\\
\end{tabular}
\end{quote}
\begin{description}
\item[method] shall be the \emph{method} used in HTTP request. As mentioned in Section \ref{http-request-response}, this shall be \verb+POST+.
\item[referer] shall be the value of Referer field in HTTP request header.
\end{description}
\subsubsection{Response}
On successful verification of signature, \emph{ssh-webagent} shall send a message containing \emph{data} of type KEX\_DH\_RESPONSE in the format described below,
\begin{quote}
\begin{tabular}{l p{1cm} l}
mpint &  & f\\
string &  & E\\
\end{tabular}
\end{quote}
\begin{description}
\item[f] is the computed value of \emph{ssh-webagent} in key exchange.
\item[E] is the encrypted \emph{message body} of type NEW viewable only by \emph{trusted server}. \emph{message body} is described, in detail, in Section \ref{message-body}.
\end{description}
\subsubsection{Secret}
The secret \textbf{S} is computed by \emph{ssh-webagent} through Diffie-Hellman key exchange and it is used to compute a \emph{shared secret}, a \emph{secret key} and an \emph{initialization vector}.
\begin{enumerate}
\item{\textbf{Shared Secret}}\\
\emph{shared secret} is the computed hash of the following values using 256-bit Secure Hash Algorithm (SHA-256) \cite{sha}.
\begin{quote}
\begin{tabular}{l p{1cm} l}
string &  & method\\
string &  & referer\\
mpint &  & e\\
mpint &  & f\\
mpint &  & S\\
\end{tabular}
\end{quote}
\item{\textbf{Secret Key}}\\
\emph{secret key} shall be used as the key for symmetric encryption. A SHA-256 hash of the following values is the \emph{secret key}.
\begin{quote}
\begin{tabular}{l p{1cm} l}
mpint &  & S\\
string &  & shared secret\\
byte &  & 'A'\\
string &  & referer\\
\end{tabular}
\end{quote}
\item{\textbf{Initialization Vector}}\\
Depending on the encryption algorithm, if needed, the \emph{initialization vector} can be obtained by computing a SHA-256 hash of the following values,
\begin{quote}
\begin{tabular}{l p{1cm} l}
mpint &  & S\\
string &  & shared secret\\
byte &  & 'B'\\
string &  & referer\\
\end{tabular}
\end{quote}
\end{enumerate}
Once the \emph{trusted server} receives the response message. It computes its own values for the above and decrypts the \emph{message body}.
\subsection{Message Body}
\label{message-body}
The \emph{message body} holds the cipher text for communication between \emph{ssh-webagent} and \emph{trusted server}. It also includes the session identifier and the encryption algorithm for decryption purposes. The type of message body and its related contents are encrypted and stored in cipher text.

\subsubsection{format}
\label{message-body-format}
The binary format of message body is shown below,
\begin{quote}
\begin{tabular}{l p{1cm} l}
byte &  & algorithm\\
string &  & identifier\\
string &  & ciphertext\\
\end{tabular}
\end{quote}
\begin{description}
\item[algorithm] used for encryption can be identified using this field. Currently supported encryption algorithms and their respective values are listed below,

\begin{tabular}{l p{1cm} l}
AES\_256\_CBC && 0x02\\
\end{tabular}
\item[identifier] is the session identifier.
\item[ciphertext] is the encrypted part of message body.
\end{description}
\subsubsection{ciphertext and plaintext}
The encrypted contents of \emph{ciphertext}, namely the \mbox{\emph{plaintext}}, shall be in the format specified below,
\begin{quote}
\begin{tabular}{l p{1cm} l}
byte[4] &  & random\\
byte &  & type\\
string &  & identifier\\
\multicolumn{1}{c}{\ldots} & \multicolumn{1}{c}{\emph{payload}} & \multicolumn{1}{c}{\ldots}\\
byte[n] &  & padding\\
\end{tabular}
\end{quote}
\begin{description}
\item[random] shall be a 32-bit (4 bytes) cryptographically secure random number.
\item[type] indicates the message body type and it shall be used to interpret the \emph{payload}. The supported types are listed below,

\begin{tabular}{l p{1cm} l}
NEW && 0x02 \\
AUTH\_REQUEST && 0x03 \\
AUTH\_RESPONSE && 0x04 \\
\end{tabular}
\item[identifier] is the session identifier and it shall be the same as \emph{identifier} defined in Section \ref{message-body-format}.
\item[payload] shall be zero or more bytes of \emph{message} content. The format of \emph{payload} depends on message body type and they are described in detail in their respective sections.
\item[padding] shall be zero or more bytes of random padding to meet the block size requirements of encryption algorithms.
\end{description}
\subsubsection{NEW}
\label{message-body-NEW}
First \emph{message} with a message body, in a \emph{session}, is always of type NEW. This message is sent by \emph{ssh-webagent} to \emph{trusted server}. The format of \mbox{\emph{plaintext}} in message body of type NEW is shown below,
\begin{quote}
\begin{tabular}{l p{1cm} l}
byte[4] &  & random\\
byte &  & type\\
string &  & identifier\\
byte[n] &  & padding\\
\end{tabular}
\end{quote}
It should be noted that \emph{payload} is \emph{empty} and, hence, is not shown in \emph{plaintext}.
\subsection{Authentication}
The authentication process shall be initiated by \emph{trusted server} after establishing a \emph{session}. The messages used in the process are as defined in Section \ref{message} and shall contain \emph{message body} as \emph{data} and shall use \emph{type} PRIVATE.
\subsubsection{Request}
A request for ``publickey'' authentication is sent by the \emph{trusted server} in the form of a \emph{message} containing \emph{data} of type PRIVATE and \emph{message body} of type AUTH\_REQUEST.
\\
This request provides the secret SSH \verb+session+ \verb+identifier+\footnote{session identifier as shown in SSH message, Section \ref{sshmsg}.} required to generate SSH message for signing and authentication. The format of the request \emph{message} is shown below,
\begin{quote}
\begin{tabular}{l p{1cm} l}
string &  & ``SSHWebAgent''\\
byte &  & VERSION\_1\_1\\
byte &  & PRIVATE\\
string &  & \emph{message body}\\
\end{tabular}
\end{quote}
The format of \emph{message body} is as follows,
\begin{quote}
\begin{tabular}{l p{1cm} l}
byte &  & algorithm\\
string &  & identifier\\
string &  & \emph{ciphertext}\\
\end{tabular}
\end{quote}
And, the format of \emph{plaintext} that is encrypted to form the \emph{ciphertext} is,
\begin{quote}
\begin{tabular}{l p{1cm} l}
byte[4] &  & random\\
byte &  & AUTH\_REQUEST\\
string &  & identifier\\
string &  & SSH \verb+session identifier+\\
byte[n] &  & padding\\
\end{tabular}
\end{quote}
SSH \verb+session+ \verb+identifier+ is as defined in RFC 4253 \cite{rfc4253}, Section 7.2, ``Output from Key Exchange''
\subsubsection{Response}
In response to the request, \emph{ssh-webagent} shall use SSH \verb+session+ \verb+identifier+, construct the SSH message and sign it with one or more private keys of the user. The resulting signatures are sent securely in a response \emph{message} to the \emph{trusted server} for successful authentication. The format of \emph{plaintext} in response \emph{message} is as follows,
\begin{quote}
\begin{tabular}{l p{1cm} l}
byte[4] &  & random\\
byte &  & AUTH\_RESPONSE\\
string &  & identifier\\
boolean &  & status\\
string &  & \emph{signatures}\\
string &  & \emph{options}\\
byte[n] &  & padding\\
\end{tabular}
\end{quote}
\begin{description}
\item[status] shall indicate whether the signing process was successful or not. It shall be used to communicate failures to \emph{trusted server}.
\item[signatures] contain one or more signatures of the SSH message, along with the corresponding public key and comments, if any. They are formatted as follows,
\begin{quote}
\begin{tabular}{p{2cm} p{.1cm} l}
uint32 &  & \emph{n}\\
string[\emph{n}] &  & \emph{item}\\
\end{tabular}
\end{quote}
And, each \emph{item} shall use the format shown below,
\begin{quote}
\begin{tabular}{p{2cm} p{.1cm} l}
string &  & publickey\\
string &  & signature\\
string &  & comment\\
\end{tabular}
\end{quote}
\emph{string[n]} shall represent multiple strings, with the number of strings indicated by \emph{n}. When \emph{n} is zero, \emph{string[n]} shall be empty and, hence, shall not be present.
\item[options] shall be used to pass information to the \emph{trusted server} in the form of key-value pairs. It shall use the format as specified below,
\begin{quote}
\begin{tabular}{p{2cm} p{.1cm} l}
uint32 &  & \emph{n}\\
byte &  & es\\
string[\emph{n}] &  & \emph{option}\\
\end{tabular}
\end{quote}
And, each \emph{option} shall hold a \emph{key} and a \emph{value} as shown below,
\begin{quote}
\begin{tabular}{p{2cm} p{.1cm} l}
string &  & key\\
string &  & value\\
\end{tabular}
\end{quote}
The \emph{value} shall be encrypted using an encryption scheme \emph{es}. The following encryption schemes shall be supported,
\begin{quote}
\begin{tabular}{l p{0.5cm} l}
PKCS1\_RSAES\_OAEP &  & 0x02\\
\end{tabular}
\end{quote}
PKCS1\_RSAES\_OAEP encryption scheme shall use the public key of \emph{trusted server} to encrypt the \emph{value}.
\end{description}
On receiving the response message, the \emph{trusted server} shall have all that it needs to perform ``publickey'' authentication in Secure Shell (SSH) protocol.
\section{Performance}
The authentication method ensures that the remote login to server is secure. But, for the task at hand, it is most likely a hindrence and sooner it is done, the better. The primary contributing factors to longer authentication time would be network latency to trusted server. Data encryption, signing and verification are reasonably quick on modern processors and are negligible in comparison. The latency to web agent will also be negligible since it resides on localhost.

The protocol uses HTTP requests to receive and send messages. Hence, session establishment and authentication process will each require three HTTP requests, when done synchronously. First request shall be used to get the message (request) from trusted server, second request shall be used to pass the message (request) to web agent and receive a message (response) and the final request shall be used to pass the message (response) to trusted server. With two requests to trusted server per session establishment and authentication process. A total of four requests to trusted server remain the primary contributing factors to performance.

HTTP request to trusted server will be over TLS. Hence, apart from the usual TCP connection negotiation, a TLS session negotiation will also happen when a new connection is established. So, the first request will take a significantly longer time, in comparison to subsequent requests to trusted server. The total time taken to complete four requests can be significant and can be a hinderance. To achieve better performance, it is recommended that the network latency to trusted server remains below 80ms.
\section{Conclusion}
Secure Shell protocol has been commonly used for remote login to servers. It is the first step to managing servers remotely. As the number of servers increase, the problem of logging to multiple servers, a time consuming task, become apparent. These problems can be solved to an extent through applications which abstract out and automatically login to servers. However, these applications may require the private key of user for automatic login.

In a world with increasing use of virtualization, more and more servers (or virtual servers) are purchased from providers and managed remotely over public internet using a web browser. In the absense of a method for secure remote login to these servers, users are left to use traditional tools for login or at worst, take a security risk and, give away the private key to third-party applications.

Hence, solving the problem of secure remote login (``publickey'') over the web, ie., HTTP, will open doors for easier access to servers. And, bring back private keys to the hands of the user. When combined with a powerful server management application, this can bring servers to general public, who, without any technical knowledge, can avail internet services through their servers in a decentralized manner.

\newpage
\appendix
\section{Trusted Servers File}
\label{trusted-servers-file-format}
The public key\footnote{Public key should be in a format as defined in RFC 4253 \cite{rfc4253}, Section 6.6 ``Public Key Algorithms''} of trusted server and its acceptable HTTP Referer \underline{prefixes} are the contents of this file. Each trusted server has an entry starting with its public key on the first line, followed by one or more acceptable prefixes of HTTP Referer values per line. The entry for trusted server ends with a line containing a single dot (.), followed by an entry for next trusted server.

A sample trusted servers file in this format has been shown below,

\begin{tabular}{l|p{3cm}}
1 & \verb|AAAAB3NzaC1yc2EAAAADAQABAAABA|
\verb|QCrLJjgFEA7tLyydh5eS2DCglbS5/|
\verb|767i5MaCoXoxZAphI/JqTD62nYJ6P|
\verb|G0hYu8spE2kcTtNHl0mcsygFEaa8v|
\verb|lFjxYL7dW/HuOfayQ+eHZq/xPtTlu|
\verb|oSOW6yI9qKj1fnaxF9IHtdZVOkcSw|
\verb|uEmlJfKjogf6Nbyn8M+biMK6Py5K4|
\verb|sll0sN48WGYEz9Xe8CcZJdhCyw97f|
\verb|hPELlXwCqvQjGqXpekSWTe4lpiQKv|
\verb|1Zn6T7/E5VW0mvu419WkLlAU7qZ1x|
\verb|fW5bfqggXnGnmOSawRGWzFaOEtsHJ|
\verb|Wn4lc//OHiWYj0MRkLd7+VVsBEF+O|
\verb|C2IAzJ4QyQtlecLkmu5Yfq/Z|
\\
2 & \verb|https://webssh.example.com/ssh/|\\
3 & \verb|https://webssh.example.com:444/ssh/|\\
4 & \verb|https://webssh.example.com:445/ssh/|\\
5 & \verb|.|\\
6 & \verb|AAAAB3NzaC1yc2EAAAADAQABAAABA|
\verb|QDPyjl9euMQ4Crj/0VyP69+ltELAM|
\verb|4Wt0GyG8y3ENEtpa/Qv0XcJ1IZ8l3|
\verb|lRRWt5+ame2LKQJwInK1xo3UqL+Jd|
\verb|CA1OX9h1ap8wOWEm6ZHiehB0JNe7B|
\verb|gIwPYl69qLpv48Xywtz28BahxZPSD|
\verb|d7k5NxiH4HIUbau3tHlvsO2LOqj9p|
\verb|QOPEDh+GdmMcgEv0ZQMY9B6uKJqI+|
\verb|RdiDgWHNDUW+pFwRi2xzMFQqPCqC0|
\verb|7ykKMI8G/Nl3Q7RQuDiRw9AhO/Brd|
\verb|F1NEa3I4fyg09nPkBP351kBrLl17V|
\verb|PgoVP24VZJkZSojEKnp4KkIhGLTfg|
\verb|+5TqI6kx36blHZpx3g8txAQt|
\\
7 & \verb|https://sshclient.example.com/|\\
8 & \verb|.|\\
\end{tabular}
\section{Local IP, Port and Domain}
\label{local-ip-port}
SSH Web Agent binds to local loopback address and accepts request on localhost. The IPv4 address range assigned for local loopback is 127.0.0.0/8.\footnote{RFC 6890, ``Special-Purpose IP Address Registries''} Any address in this range can be used by web agent for accepting requests. However, for easier inter-operability, the IPv4 address, in this range, \textbf{127.82.11.29} and port number \textbf{8211} is suggested for use.

HTTP/TLS protocol requires web agent to use a TLS certificate signed by a trusted certificate authority. TLS certificates are, typically, issued for a domain instead of an IP address.

TLS certificate and its private key will need to be distributed along with web agent. Hence, they should be considered public. And, on the same note, for security reasons, it is recommended that the entire domain is considered public and subdomains are not used for any other purpose.

The domain \textbf{localhost-ssh-webagent.in} has been assigned for this purpose and it resolves to the suggested IPv4 address, 127.82.11.29. The \emph{whois} information for localhost-ssh-webagent.in provides the current contact to obtain TLS certificates.
\section{Connection Owner}
\label{connection-owner}
A web agent listening on loopback interface can be connected to by any user. This
results in a privilege escalation and unauthorized access to private key. This security issue can be
mitigated by ensuring that the remote end of the local connection belongs to the owner of the web agent
process itself. The premise here is that the owner of the web agent process is given access to the private key
and hence can allow access to the same user.

On linux kernels, \verb+/proc/net/tcp+ provides a formatted output of active connections. In this output, the columns \emph{local\_address}, \emph{rem\_address} and \emph{uid} are of relevance to us. The IPv4 address and port in the output are shown in hexadecimal. 

The sample output below shows the web agent listening on 127.82.11.29:8211 (1), a SSH server listening on 0.0.0.0:22 (2) and a SMTP server listening on 127.0.0.1:25 (3). It should be noted that the \emph{uid} for web agent is 1000, indicating a user, while the \emph{uid} for others is, 0 indicating the superuser.
\begin{quote}
\begin{tabular}{l|l}
 & \verb|local_address |\verb| rem_address  |\verb|uid|\\
1 & \verb|1D0B527F:2013 00000000:0000 1000| \\
2 & \verb|00000000:0016 00000000:0000 0| \\
3 & \verb|0100007F:0019 00000000:0000 0| \\
\end{tabular}
\end{quote}

The sample output below shows active local connections,
\\
\\
\begin{tabular}{l|l}
\verb|local_address |\verb| rem_address  |\verb|uid| & \\
\verb|1D0B527F:2013 00000000:0000 1000| & listener\\
\verb|1D0B527F:2013 0100007F:CE93 1000| & accept\\
\verb|1D0B527F:2013 0100007F:CE94 0| & \emph{reject}\\
\verb|0100007F:CE93 1D0B527F:2013 1000| & -\\
\verb|0100007F:CE94 1D0B527F:2013 0| & -\\
\end{tabular}
\\
\par Using the output above, web agent can easily compare the \emph{uid} of connection with the process \emph{uid}. In case of a mismatch, the connection has originated from a different user and should be rejected.
\end{document}